\documentclass{article}

\usepackage{arxiv}

\usepackage[utf8]{inputenc} 
\usepackage[T1]{fontenc}    
\usepackage{hyperref}       
\usepackage{url}            
\usepackage{booktabs}       
\usepackage{amsfonts}       
\usepackage{nicefrac}       
\usepackage{microtype}      
\usepackage{lipsum}
\usepackage{graphicx}
\usepackage{float}

\usepackage[numbers,sort&compress]{natbib} 
\usepackage{amssymb,amsthm,amsmath}

\title{Rules create unequal rewards: Elite tennis players allocate resources efficiently}

\author{
  Masatsugu Yoshizawa \\
  Department of Life Sciences (Sports Sciences)\\
  Graduate School of Arts and Sciences\\
  The University of Tokyo\\
  Meguro-ku, Tokyo 153-8902, Japan \\
  \And
  Yuta Kawamoto \\
  The University of Tokyo Sports Science Initiative (UTSSI)\\
  The University of Tokyo\\
  Meguro-ku, Tokyo 153-8902, Japan \\
  \And
  Daisuke Takeshita\thanks{Corresponding author.} \\
  Department of Life Sciences (Sports Sciences)\\
  Graduate School of Arts and Sciences\\
  The University of Tokyo\\
  Meguro-ku, Tokyo 153-8902, Japan \\
  \texttt{dtakeshita@idaten.c.u-tokyo.ac.jp} \\
}

\begin{document}
\maketitle

\begin{abstract}
In many competitive settings, from education to politics, rules do not reward effort evenly, and thresholds (e.g., grade cutoffs or electoral majorities) make some moments disproportionately important. Success thus depends on efficiently allocating limited resources. However, empirical demonstration has been difficult because effort allocation is rarely observable and feedback is often delayed, limiting our understanding of expertise. Professional tennis provides an ideal natural experiment. Because each game resets after a player wins four points and points in a lost game are wasted, the value of a point varies sharply across scores. Efficient allocation should therefore win games without wasting points, conserving resources for future games. Such allocation manifests in score-dependent point-winning probabilities, from which we derive each player’s Pareto frontier---the theoretical limit of the trade-off between game-winning probability and the expected points per game. Here, we show that top players operate closer to this frontier, converting points to game wins more efficiently. Optimal strategies reduce the probability of winning points when the player is far behind (e.g., 0--2, 0--3). This behavior is psychologically difficult---letting go of the current game---but represents a rational energy conservation strategy. Top players exhibit this pattern especially in return games, where winning points is harder than in service games, requiring them to drastically vary their efforts, consistent with game-theoretic predictions. These findings suggest that elite performance reflects efficient adaptation to rule-created value structures; knowing when to give up may be as fundamental to expertise as knowing when to compete.
\end{abstract}


\newpage
\section{Introduction}
In most sports, the player with more points wins. Tennis is different. In the 2019 Wimbledon men's final, Roger Federer won 14 more total points than Novak Djokovic (218 vs. 204) but lost the match. This paradox arises from the sport's scoring rule. Unlike sports such as soccer or basketball, where total points determine the winner, tennis---like volleyball and badminton---requires winning games and sets. Each game typically resets after a player wins four points, meaning that points won in a lost game do not carry over and are wasted \citep{Lisi2019, Wright2013}. This threshold-based, winner-take-all structure at the game level creates variation in point values by score, making \textit{when you win points} more important than \textit{how many you win} \citep{Barnett2004, Croucher1986, Klaassen2001, Morris1977, Odonoghue2001}. Allocating energy to these wasted moments is inefficient; therefore, tennis can be viewed as a resource optimization problem: deciding at which scores to invest limited resources---effort and focus.

Rules that create unequal returns to effort are not unique to sports. In many other competitive domains, thresholds make certain moments disproportionately important, shaping how people allocate resources. For instance, students focus their efforts on meeting grade cutoffs \citep{Attali2016, Gibbs2004} and researchers focus their efforts on publishing in top journals that yield greater career rewards \citep{Siler2015, Trueblood2025}. Recent research shows that such reward structures can influence decision-making and behavior, inducing stress or unethical conduct \citep{Hu2024, McAuliffe2022, Murtaza2023}. However, much of this work relies on controlled experiments or subjective evaluations, leaving a gap in our understanding of how efficiently people adapt to unequal reward structures in high-stakes real-world settings. Because, in such domains, effort is often unobservable and feedback is delayed, it is difficult to link moment-to-moment decisions to outcomes.

Professional tennis provides an ideal natural experiment to address this gap. The sport generates abundant point-level data with immediate outcomes and high career stakes, which ensure motivation \citep{Gossner2012, Klaassen2001, PrietoLage2023, Walker2001}. This enables observation of rational behaviors under high-stakes conditions that are difficult to replicate in laboratory settings \citep{Gauriot2016, Walker2001}. Moreover, the variation in point value \citep{Barnett2004, Croucher1986, Klaassen2001, Morris1977, Odonoghue2001} requires players to allocate resources across scores. Such allocation manifests in score-dependent point-winning probabilities, acting as behavioral markers of effort and focus. Given the four-point threshold for winning games, efficient allocation should yield more game wins with fewer points played---conserving resources for subsequent games. However, previous studies either assume a constant point-winning probability across all scores \citep{Newton2005, OMalley2008, Pollard1983} or incorporate score-dependence without evaluating efficient adaptation to the game's value structure \citep{Carrari2017, Ferrante2017, Sim2020}.

Here, we show that top players convert points into game wins more efficiently through optimized score-dependent point-winning probabilities. We model resource allocation by treating each player's average point-winning probability as a resource constraint representing their ability to win points. Under this constraint, we derive the Pareto frontier---the theoretical limit linking game-winning probability and expected points per game---and quantify efficiency as the distance to this frontier. Elite players perform closer to this limit, indicating superior adaptation to the game's reward structure. By measuring this adaptation where theoretical optima and actual behavior are observable, our findings provide a quantitative basis for assessing strategic efficiency under threshold-based incentive structures in other domains.
\newpage
\section{Results}
\begin{figure}[H]
  \centering
  \includegraphics[width=1.0\linewidth]{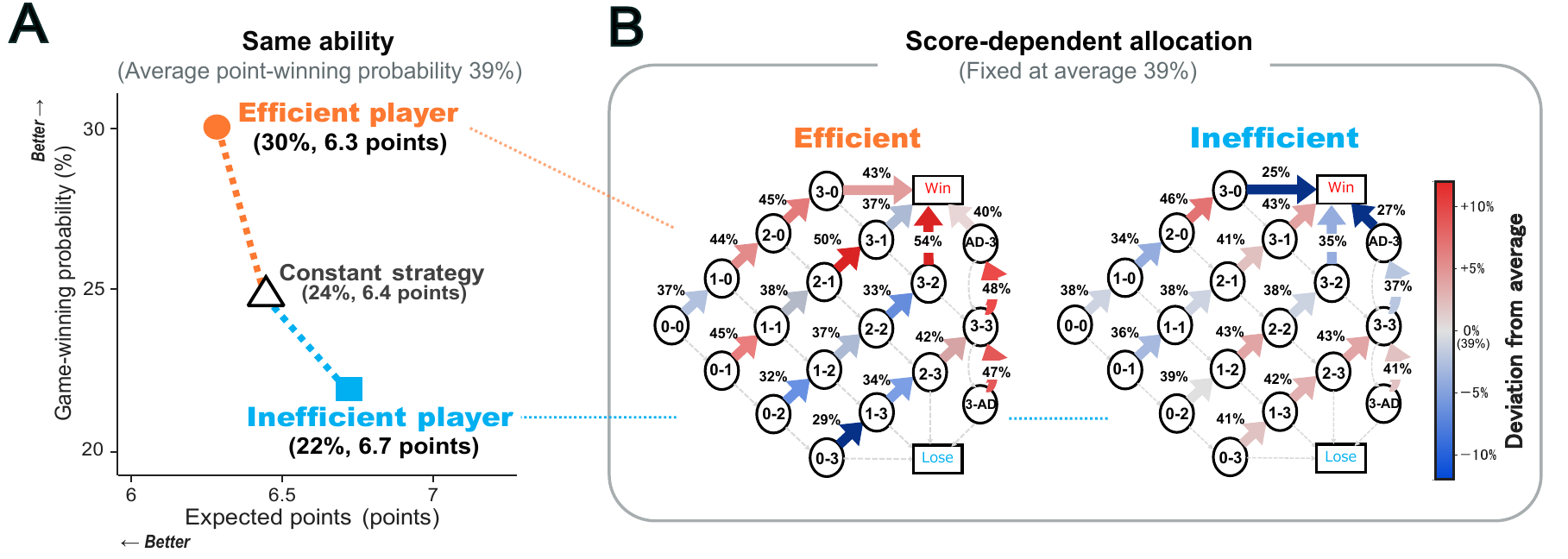}
  \caption{\textbf{Same ability, different outcomes.}
  \textbf{(A) Two players with the same average point-winning probability (39\%) but different results.} This example from women's return games illustrates how the efficient player achieves a higher game-winning probability with fewer expected points per game compared to the inefficient player or a constant-strategy baseline (triangle; fixed 39\% at all scores). \textbf{(B) Score-dependent allocation.} Scores reflect the players' position within each game (e.g., 2--0 means the player has 2 points, opponent has 0). The efficient player shows higher probabilities at leading or neutral scores (e.g., 45\% at 2--0, 50\% at 2--1) and lower probabilities at trailing scores (e.g., 29\% at 0--3), whereas the inefficient player shows the opposite pattern. Tennis games require winning at least four points with a two-point margin; at 3--3 (deuce), play enters the advantage (AD) cycle. Tennis scoring (0, 15, 30, 40, AD) presented using simplified notation (0, 1, 2, 3, AD) for clarity.}
  \label{fig:fig1}
\end{figure}

\textbf{An example: Same ability, different outcomes.} We analyzed service and return games separately because serving provides a large advantage in average point-winning probability (men: 64\% vs. 37\%, women: 58\% vs. 44\%). Figure~\ref{fig:fig1}A shows two players from our women's return dataset with the same average point-winning probability (39\%) but contrasting outcomes in game-winning probability and expected points per game. The efficient player won 30\% of the return games and required 6.3 points per game, converting the same ability into more game wins with fewer points. By contrast, the inefficient player won only 22\% of games and needed 6.7 points per game, playing more points yet winning fewer games. 

Figure~\ref{fig:fig1}B shows that the efficient player exhibits greater point-winning probabilities at leading scores (e.g., 60\% at 2--1, 64\% at 3--2) and lower probabilities at trailing scores (e.g., 32\% at 0--2, 29\% at 0--3), whereas the inefficient player shows the opposite pattern. We observe variation across players in score-dependent probability allocation (SI Appendix, Fig.~S1), suggesting that such variation may underlie the differences in efficiency.

\newpage
\textbf{Statistical superiority of the score-dependent model.} To evaluate whether score-dependent strategies better fit player-level game outcomes, we compare two models for each player. The \textit{constant model} assumes a uniform point-winning probability across all 18 scores, whereas the \textit{score-dependent model} allows probabilities to vary across these scores. We evaluate model fit using the Akaike information criterion (AIC)~\citep{Akaike1974}, Bayesian information criterion (BIC)~\citep{Schwarz1978}, and adjusted $R^2$. Across 226{,}795 games from 178 players (84 men and 94 women), the score-dependent model provides a better fit than the constant model for nearly all metrics (Table~\ref{tab:model_comparison}). Although BIC favors the simpler model for women's service, the other metrics support the superiority of the score-dependent model. These results show that the score-dependent model captures the structure of players’ score-dependent behavior, providing a robust basis for the subsequent efficiency analyses.

\begin{table}[H]
\caption{Statistical superiority of the score-dependent model}
\label{tab:model_comparison}
\centering
\setlength{\abovecaptionskip}{5pt}  
\setlength{\belowcaptionskip}{2pt}   
\renewcommand{\arraystretch}{0.9} 
\small                                
\begin{tabular}{lrrr}
\toprule
\multicolumn{1}{c}{Category / Metric} &
\multicolumn{1}{c}{Const.} &
\multicolumn{1}{c}{Score-dep.} &
\multicolumn{1}{c}{Difference$^{a}$} \\
\midrule
\multicolumn{4}{l}{\textbf{Game-winning probability}}\\[-2pt]
Men's service (AIC)         & -836.36 & \textbf{-925.38}  & -89.02 \\
Men's service (BIC)         & -833.78 & \textbf{-878.85}  & -45.07 \\
Men's service (Adj.$R^2$)   &  0.942  & \textbf{0.980}    & -- \\[2pt]
Men's return (AIC)          & -770.82 & \textbf{-1172.96} & -402.14 \\
Men's return (BIC)          & -768.24 & \textbf{-1126.43} & -358.19 \\
Men's return (Adj.$R^2$)    &  0.863  & \textbf{0.998}    & -- \\[2pt]
Women's service (AIC)       & -953.26 & \textbf{-955.71}  &  -2.45 \\
Women's service (BIC)       & \textbf{-950.58} & -907.43  & +43.15 \\
Women's service (Adj.$R^2$) &  0.964  & \textbf{0.969}    & -- \\[2pt]
Women's return (AIC)        & -894.34 & \textbf{-1083.62} & -189.28 \\
Women's return (BIC)        & -891.66 & \textbf{-1035.35} & -143.69 \\
Women's return (Adj.$R^2$)  & 0.944   & \textbf{0.992}    & -- \\
\midrule
\multicolumn{4}{l}{\textbf{Expected points per game}}\\[-2pt]
Men's service (AIC)         & -401.67 & \textbf{-565.05}  & -163.38 \\
Men's service (BIC)         & -399.08 & \textbf{-518.53}  & -119.45 \\
Men's service (Adj.$R^2$)   &  0.575  & \textbf{0.931}    & -- \\[2pt]
Men's return (AIC)          & -391.00 & \textbf{-555.98}  & -164.98 \\
Men's return (BIC)          & -388.42 & \textbf{-509.45}  & -121.03 \\
Men's return (Adj.$R^2$)    &  0.503  & \textbf{0.921}    & -- \\[2pt]
Women's service (AIC)       & -414.59 & \textbf{-505.53}  &  -90.94 \\
Women's service (BIC)       & -411.91 & \textbf{-457.25}  &  -45.34 \\
Women's service (Adj.$R^2$) &  0.247  & \textbf{0.718}    & -- \\[2pt]
Women's return (AIC)        & -364.73 & \textbf{-512.80}  & -148.07 \\
Women's return (BIC)        & -362.05 & \textbf{-464.52}  & -102.47 \\
Women's return (Adj.$R^2$)  & -0.011  & \textbf{0.777}    & -- \\
\bottomrule
\end{tabular}
\renewcommand{\arraystretch}{1.0}    

\vspace{3pt}
\raggedright
\footnotesize
Values are averaged across players (men: $n=84$; women: $n=94$). Bold figures indicate a better fit. Const.: constant model assuming a fixed point-winning probability across all scores. Score-dep.: score-dependent model allowing probabilities to vary across 18 scores. $^{a}$Difference = score-dependent minus constant. Negative values favor the score-dependent (AIC/BIC). 
\end{table}

\begin{figure}[H]
\centering
\includegraphics[width=.85\linewidth]{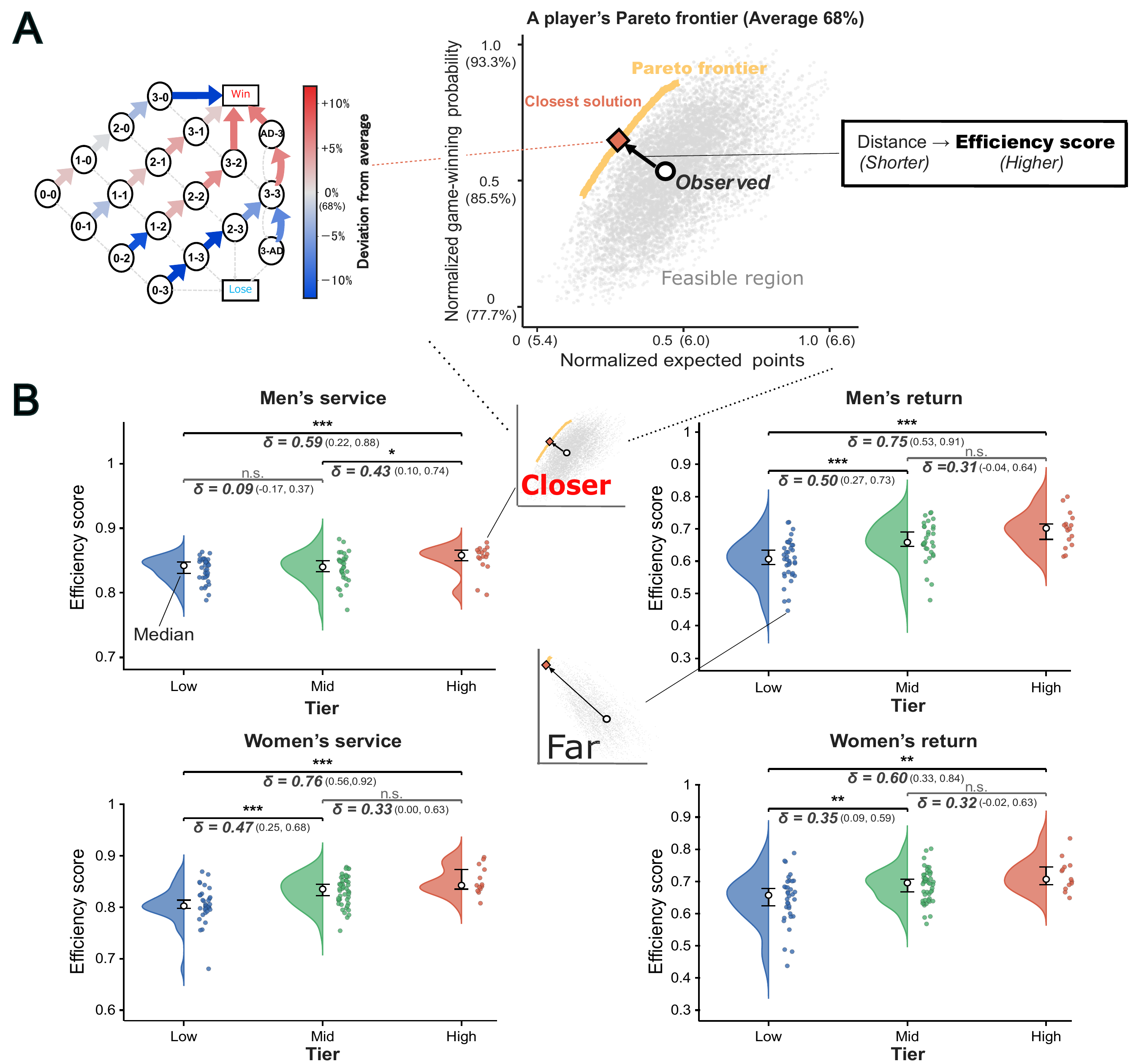}
\caption{\textbf{Elite players convert points into game wins more efficiently.} 
\textbf{(A) Quantifying efficiency.} Using an example player, the efficiency score quantifies proximity to the theoretical Pareto frontier (yellow curve) in the normalized outcome space. The score is derived from the shortest Euclidean distance between the observed performance and the frontier scaled to a 0--1 range, where 1 represents the theoretical optimum (orange diamond).
\textbf{(B) High-tier players show higher efficiency scores.} Efficiency scores are compared for service and return games across three match-winning tiers: low ($<50\%$; men $n=37$, women $n=34$), mid (50--70\%; men $n=31$, women $n=47$), and high ($>70\%$; men $n=16$, women $n=13$). The white circles indicate medians, while the error bars show 95\% CIs. The horizontal bars denote Bonferroni-corrected Mann--Whitney comparisons ($*$ $p<0.05$, $**$ $p<0.01$, $***$ $p<0.001$, n.s.: not significant) with effect sizes (Cliff’s $\delta$).}
\label{fig:fig2}
\end{figure}

\newpage
\textbf{Elite players convert points into game wins more efficiently.} Using the score-dependent model, we quantify strategy efficiency on a 0--1 scale (1 = theoretical optimum). This metric reflects the proximity of a player's observed performance to their theoretical Pareto frontier (Fig.~\ref{fig:fig2}A), defined as the optimal trade-off between game-winning probability and the expected points per game, constrained by their average point-winning probability. The frontier represents the theoretical efficiency boundary for which a player maximizes their game outcomes for a given skill level. We compare efficiency scores across the match-winning tiers in Grand Slam tournaments: high ($>$70\%, expected two or more wins), middle (50--70\%, 1--2 wins), and low ($<$50\%, fewer than one win). Because normality assumptions are violated for some groups (Lilliefors test, $p < 0.05$), we use Cliff’s delta ($\delta$) for pairwise comparisons. High-tier players outperform low-tier players in men's service ($\delta = 0.59$, 95\% confidence interval (CI) [0.24, 0.88], $p < 0.001$) and return games ($\delta = 0.75$, 95\% CI [0.53, 0.91], $p < 0.001$), as well as women's service ($\delta = 0.76$, 95\% CI [0.55, 0.93], $p < 0.001$) and return games ($\delta = 0.60$, 95\% CI [0.36, 0.83], $p = 0.002$). These results show that elite players operate closer to their theoretical efficiency limits, highlighting efficiency as a key dimension of competitive performance.

\begin{figure}[H]
\centering
\includegraphics[width=1.0\linewidth]{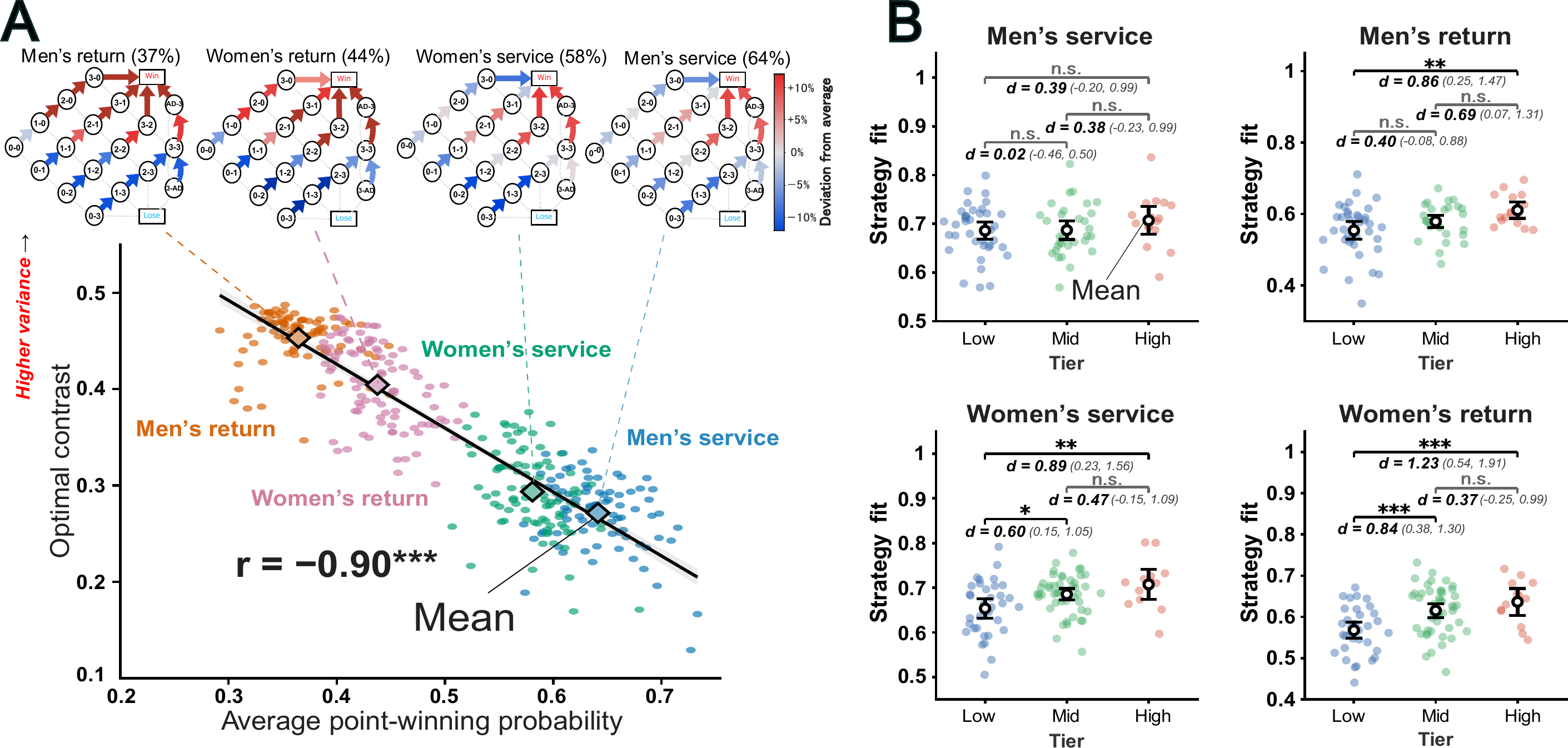}
\caption{\textbf{Optimal allocation patterns and strategy fit.}
\textbf{(A) Lower baseline ability requires higher strategic contrast.} A strong negative correlation ($r = -0.90$) links lower average point-winning probabilities to higher optimal contrast, indicating that weaker positions (e.g., return games) demand more extreme strategies to maximize outcomes. Points represent individual players (men $n=84$, women $n=94$); diamonds denote category means; the black line shows linear regression with 95\% CIs. Upper state diagrams illustrate optimal patterns; arrows indicate probability deviation from the category mean. 
\textbf{(B) High-tier players follow the optimal pattern.} Strategy fit---quantifying the proximity of actual allocation to theoretical optima in 18-dimensional space--- is compared across the low ($<50\%$), middle ($50$–$70\%$), and high ($>70\%$) match-winning tiers. Dots represent players; white circles show tier means with 95\% CIs. Horizontal bars indicate pairwise comparisons with Cohen’s $d$ (* $p<0.05$, ** $p<0.01$, *** $p<0.001$, n.s.: not significant).}
\label{fig:fig3}
\end{figure}

\textbf{Optimal allocation patterns and strategy fit.} Players’ allocation patterns reflect the structural logic of tennis scoring, that is, when baseline success rates are low, maximizing outcomes involves stronger strategic contrasts (i.e., larger variance in effort across scores). This principle is quantified in the scatter plot in Fig.~\ref{fig:fig3}A, which shows a strong negative correlation between average point-winning probability and optimal contrast ($r = -0.90$, $p < 0.001$). Consequently, the same player optimally adopts different strategies depending on their role, as operating from a weaker position (return games) demands stronger contrasts than operating from a stronger one (service games).

Specific allocation patterns are illustrated in the state diagrams in Fig.~\ref{fig:fig3}A. In service games, where average point-winning probabilities are high (men 64\%, women 58\%), optimal strategies concentrate resources on close, late-game scores (e.g., 2--2, 3--2). By contrast, return games have lower average probabilities (men 37\%, women 44\%). The theoretical optima call for consistently elevated probabilities across break opportunities (2--0 to 3--2, AD--3), combined with sharper reductions at trailing scores (e.g., 0--2, 0--3). 

We then examine whether elite players approximate these optimal patterns. We quantify "strategy fit" as a scaled score (from 0 to 1) measuring the proximity of a player’s observed allocation to their closest theoretical Pareto-optimal solution (see Materials and Methods). Higher values represent a closer alignment with the optimal pattern. As shown in Fig.~\ref{fig:fig3}B, strategy fit consistently increases with match-winning tiers. Women's return games show the strongest hierarchy, with large effects distinguishing the high ($>70\%$ match win rate) from the low ($<50\%$) tiers ($d = 1.23$, 95\% CI [0.54, 1.91]). Similar patterns are seen for women's service ($d = 0.89$, [0.23, 1.56]) and men's return games (high vs.\ low: $d = 0.86$, [0.25, 1.47]). Although men’s service games do not show statistically significant differences, the observed means follow the same hierarchy (high $>$ middle $>$ low). These results indicate that high-tier players align their resource allocation more closely with theoretical optima.

\section{Discussion}
We found that elite players convert points to game wins more efficiently, reflecting performance closer to their theoretical limits (Fig.~\ref{fig:fig2}). Unlike previous studies that examined game-winning probabilities and expected points per game as separate outcomes \citep{Carrari2017, Ferrante2017, Newton2005, OMalley2008, Pollard1983, Sim2020}, we evaluated allocation efficiency. We derived each player's Pareto frontier via multi-objective optimization using a score-dependent model constrained by average point-winning probability (Table~\ref{tab:model_comparison}, Fig.~\ref{fig:fig2}A). Efficiency, quantified as the proximity to this frontier, increases with match-winning tiers (Fig.~\ref{fig:fig2}B). Theoretically optimal strategies increase the point-winning probabilities at neutral or leading scores and reduce them when trailing (Fig.~\ref{fig:fig3}A). This pattern---stronger allocation contrasts in weaker positions (e.g., return games) ($r = -0.90$, Fig.~\ref{fig:fig3}A)---aligns with the theoretical prediction that resource constraints favor extreme strategies \citep{Chowdhury2013, Grafen1987, Morrell2005, Roberson2006, Ryvkin2022}. Overall, players with higher match-winning percentages tend to follow these optimal patterns more closely (Fig.~\ref{fig:fig3}B). These results suggest that elite performance relies not merely on point-winning ability, but also on efficient adaptation to the uneven reward structure imposed by the rules.

\textbf{Efficient play is physically smart, psychologically hard.} Elite players show a clear pattern (i.e., ``secure winnable games, do not chase likely losses'') well aligned with the four-point, two-point margin rule in tennis. They raise point-winning probabilities in neutral and leading scores and reduce them when far behind (Fig.~\ref{fig:fig3}). High-leverage situations such as deuce (3--3) and advantage (3--AD, AD--3), where a single point shifts a game’s direction, lead players to reduce risk, often producing longer rallies with higher physical demands \citep{Kovacs2006, Croucher1986, Morris1977, Odonoghue2001, Paserman2023, Prieto2018}. Allocating limited resources to high-leverage moments while conserving resources in low-return situations is consistent with the rational optimization perspective of the match-level objective function. While expending less effort contradicts the impulse to compete for every point and conflicts with loss aversion \citep{Kahneman1979}---the natural resistance to conceding the current game---selective disengagement differs from resignation. The fact that top-tier players follow this pattern more closely than their lower-tier peers (Fig.~\ref{fig:fig3}A) is consistent with a sophisticated pacing strategy rather than with a weakness \citep{Brown2014, Fernandez2009, Hornery2007, Goossens2015, Ferrante2017}. Although behavioral data cannot isolate cognitive mechanisms, maintaining selective effort requires suppressing this drive---a demand paralleling executive functions such as inhibitory control and cognitive flexibility \citep{Vestberg2012, Ren2025, Vaughan2021}. As elite athletes often possess superior executive capabilities \citep{Bonetti2025, Ren2025, Vaughan2021, Vestberg2012}, these traits may facilitate such counterintuitive forms of strategic regulation.

These findings require addressing alternative interpretations: reduced point-winning probabilities when trailing might suggest choking \citep{Baumeister1984, Hill2010}, negative momentum \citep{isoahola1980psychological}, or a lack of persistence. However, the data are inconsistent with any of these explanations. Optimal strategies require sharper reductions in point-winning probabilities at low-return scores (Fig.~\ref{fig:fig3}A), a pattern most closely followed by top-tier players (Figs.~\ref{fig:fig2}B and~\ref{fig:fig3}B). If such declines reflect psychological collapse or maladaptive resignation, they should be most pronounced among lower-tier players who lose more often. However, we observe the opposite effect: the most successful players exhibit the strongest tendency to downregulate point-winning probability in lost causes. Furthermore, transient disruptions such as choking and negative momentum would be unlikely to persist as stable, tier-graded patterns in multi-year averages. Therefore, the evidence points to the strategic downweighting of unwinnable situations rather than performance failure, indicating systematic adaptation to the value structure of the game.

\textbf{Expertise requires knowing when to try hard.} Most prior work on expertise has focused on a single objective (i.e., how to win more points), such as game-theoretic mixed strategies for serve direction or penalty kicks \citep{Gauriot2016, Palacios2003, Walker2001}. Research on the speed–accuracy trade-off \citep{Bogacz2010, Fitts1954} typically assumes that players are already using an optimal strategy. By contrast, we applied multi-objective optimization to identify each player’s theoretical limit and evaluate how closely their actual performance approaches the ideal "efficiency" frontier (Fig.~\ref{fig:fig2}A). 

While our cross-sectional design cannot firmly establish that efficiency causes success, we separated strategic efficiency from baseline technical skill by constraining the Pareto frontier to each player's average point-winning probability (i.e., their overall ability to win points). Our analysis reveals that top-tier players operate closer to this theoretical limit than their low-tier peers (Fig.~\ref{fig:fig2}B). These findings suggest that elite performance depends not only on technical execution but also on flexible resource distribution. As technical proficiency alone does not fully explain performance variance \citep{Macnamara2014}, this allocation efficiency likely represents an independent dimension of expertise. Future work should examine how this efficiency evolves over players’ careers, how it interacts with opponent level (e.g., elite-vs-elite match-ups), and how alternative scoring systems shape this strategic component of expertise by changing which points carry the greatest leverage.

\textbf{Broader implications for social science.} The institutional logic examined here, namely that threshold-based rewards create situation-specific values, applies to settings beyond sports. In systems with ``reset'' mechanisms or ``winner-take-all'' thresholds, any effort falling short of the goal is wasted. As such, students adjust study effort around grade cutoffs \citep{Attali2016, Gibbs2004}, employees target promotion standards \citep{Lazear1981}, and political campaigns focus resources on pivotal votes \citep{lindbeck1987balanced, stromberg2008electoral}. In such systems, marginal returns to effort are uneven and spike near key thresholds. Our finding that players in weaker positions adopt higher-contrast strategies (Fig.~\ref{fig:fig3}A) provides empirical support for game-theoretic predictions that resource scarcity favors extreme variance to overcome structural disadvantages \citep{Grafen1987, Morrell2005, Roberson2006, Ryvkin2022}. This dynamic underscores foundational insights from North and Acemoglu that institutions shape behavior by defining the structure of unequal rewards \citep{Acemoglu2025, North1990}. 

These parallels expose a critical learning challenge. While tennis provides explicit point values and immediate feedback, most real-world institutions operate under opaque incentives and noisy signals \citep{Simon1955, March1991}, which can severely distort learning and adaptation through cognitive limitations and biological constraints \citep{Camerer2004}. Our results show that clear value structures enable individuals to approach theoretical efficiency limits. Because experts are better able to identify where marginal returns peak, this complexity can widen the gap between the most and least successful actors. A key step is to identify how rule clarity and reliable feedback support---or impede---this adaptation. Understanding these processes is important for explaining how efficiency and inequality emerge in settings ranging from education to organizational performance.

\section{Materials and Methods}
\label{sec:methods}
\textbf{Data.} We analyzed point-by-point data from all singles matches at the four Grand Slam tournaments (Australian Open, French Open, Wimbledon, and US Open) between 2012 and 2022 from a public repository \citep{SackmannRepo}. Matches ending in retirement or walkover were excluded. Retirements confound efficiency estimates because injury-related performance declines cannot be distinguished from strategic effort reduction, whereas walkovers provide no point-level data.

To ensure reliable parameter estimation, we included only players with 30 or more matches, which yields 178 players (84 men, 94 women). The dataset includes 3,725 men's matches and 4,374 women's matches, totaling 133,434 men's games and 93,361 women's games (with 819,444 and 605,813 points, respectively).

For each player, the point-winning probabilities were computed at each of the 18 transient states (from 0--0 to advantage) by dividing the number of points won at that score by the total number of points played at that score across all observed matches. We also calculated each player's average point-winning probability across all scores, game-winning probability, and expected points per game as averages over the full observation period. This approach captures each player's long-term strategic characteristics.

\textbf{Markov chain model of game progression.} Game progression was modeled as an absorbing Markov chain with the 18 transient states and two absorbing states (game won, game lost). We compared two models for each of four categories (men's service, men's return, women's service, women's return):  
(1) a constant-probability model assuming a fixed point-winning probability across all scores and  
(2) a score-dependent model allowing distinct probabilities for each of the 18 states.  
Model fit was evaluated using the AIC~\citep{Akaike1974} and BIC~\citep{Schwarz1978}, both of which consistently favor the score-dependent model, supporting its use in the subsequent efficiency analyses.

\textbf{Player-specific Pareto frontier estimation.} To isolate the effect of score-dependent allocation while controlling for baseline ability, we jointly optimized two competing objectives---game-winning probability and expected points per game---using the score-dependent model. The Non-dominated Sorting Genetic Algorithm II (NSGA-II) \citep{Deb2002} identified each player's Pareto frontier as the set of non-dominated solutions. 

Searches were constrained within $\pm0.005$ ($\pm0.5$\%) of each player's average point-winning probability to ensure comparability while allowing strategic flexibility. Sensitivity analyses using four alternative constraints ($\pm0.0010$ to $\pm0.0075$) confirm robust efficiency rankings (SI Appendix, Table~S2). Based on data distributions, the search ranges for point-winning probability were set to 0.50--0.75 (men's service), 0.25--0.50 (men's return), 0.40--0.70 (women's service), and 0.30--0.60 (women's return). 

The NSGA-II parameters were as follows: population size 800, maximum generations 400, function tolerance $1\times10^{-4}$, crossover rate 0.8, and Pareto fraction 0.6. To mitigate stochastic variation, we ran 30 optimizations with different random seeds and combined all non-dominated solutions into a unified frontier.

\textbf{Efficiency score and tiering.} Efficiency scores were computed by comparing each player's observed performance with their own Pareto frontier. The two-dimensional outcome space (game-winning probability, expected points per game) was normalized to [0, 1], and the shortest Euclidean distance $d_{\text{out}}$ from the observed point to the frontier was calculated. The efficiency score is defined as follows: 
\[
1 - \frac{d_{\text{out}}}{\sqrt{2}},
\]
yielding values from 0 to 1, with higher values indicating performance closer to the theoretical efficiency limit.

Players were grouped into three tiers based on match-winning percentage: low ($<$50\%, fewer than one expected win per tournament; men n=37, women n=34), middle (50--70\%, typically 1--2 wins; men n=31, women n=47), and high ($>$70\%, two or more wins, on average, reaching round of 32 or better; men n=16, women n=13). 

\textbf{Tier-wise comparison of efficiency scores.} The efficiency scores across performance tiers were compared (Fig.~\ref{fig:fig2}B) after assessing normality using the Lilliefors test. Because several groups violate normality, tier differences were assessed using pairwise Mann–Whitney tests with Bonferroni correction. Effect sizes were quantified as Cliff's delta ($\delta$) with 95\% CIs estimated by bootstrap resampling (1,000 iterations). Given the retrospective observational design and large sample size, we prioritized practical significance over statistical significance, reporting effect sizes with CIs as primary outcomes and $p$ values as supplementary information. We interpreted effect sizes using conventional guidelines, considering $|\delta| \geq 0.47$ as large and practically meaningful~\citep{Romano2006NSSE}.

\textbf{Optimal allocation patterns and contrast analysis.} For each category (men's service, women's service, women's return, men's return), we extracted optimal point-winning probabilities across all 18 score states from representative solutions (Pareto-optimal points) obtained by NSGA-II. These probabilities were averaged across all players to derive category-specific optimal allocation patterns, representing the ideal strategic tendencies for each player group (Fig.~\ref{fig:fig3}A). To quantify the strength of score-dependent allocation, we calculated the standard deviation of optimal probabilities across the 18 scores for each player, defining this as the optimal contrast. We first examined the overall relationship between average point-winning probability and optimal contrast using regression analysis pooled across all players. We then conducted category-specific linear regressions stratified by sex and service/return role (see SI Appendix, Fig.~S2 for detailed results across the four categories).

\textbf{Strategy fit analysis.} To quantify how closely each player's observed allocation matches the theoretical optimum, we computed strategy fit as the Euclidean distance between observed probability vector $\mathbf{A}=(p_1,\ldots,p_{18})$ and optimal vector $\mathbf{R}=(p_1^{*},\ldots,p_{18}^{*})$ in the 18-dimensional input space. Optimal probabilities were derived from each player's closest Pareto-optimal solution. The distance is defined as follows:
\[
d_{\text{in}}=\sqrt{\sum_{i=1}^{18}(p_i-p_i^{*})^2}.
\]
Because the feasible search range differs across categories, distances were normalized by the theoretical maximum $d_{\text{in}, \max}=\sqrt{18}\,\Delta p$, where $\Delta p$ denotes the category-specific search width determined based on empirical distributions as shown in SI Appendix, Fig.~S1 (0.25 for men's service and return, 0.30 for women's service and return). Strategy fit is then defined as:
\[
\text{Strategy fit}=1-\frac{d_{\text{in}}}{d_{\text{in}, \max}},
\]
ranging from 0 to 1, with higher values indicating closer approximation to the optimal pattern. Strategy fit was compared across performance tiers (low, mid, high) after assessing normality using the Lilliefors test. Because the distributions do not violate normality assumptions ($p>0.05$), pairwise differences were evaluated using the Tukey--Kramer post-hoc test. Effect sizes were quantified as Cohen's $d$ with 95\% CIs ~\citep{Cohen1988}, with conventional thresholds for interpretation (small: $|d|\ge0.2$, medium: $|d|\ge0.5$, large: $|d|\ge0.8$).

\textbf{Study design and limitations.} This retrospective observational study uses player-level averages during 2012–2022 to estimate stable long-term strategies. The opponent-tier analysis shows no major bias in opponent strength across tiers (SI Appendix, Table~S3). Score-dependent patterns are thus considered to reflect stable strategic tendencies rather than short-term fluctuations. Nonetheless, unobserved factors, such as fatigue, psychological pressure, or surface-specific effects, could still influence point-level performance. Because of the observational design, causal inference is not possible, and the results represent associations rather than causal effects. Despite these limitations, the large dataset, balanced opponent structure, and the consistent results across genders and game types support the robustness of our findings. Future research should employ longitudinal or experimental designs to clarify how efficiency develops and to test causal mechanisms directly.

\textbf{Data Availability.} All analysis code and processed datasets are available in a public GitHub repository at \url{https://github.com/masa-yo1/EliteTennis_ResourceAllocation} and are archived in Zenodo (\url{https://doi.org/10.5281/zenodo.18266752}). The raw point-by-point data can be accessed at \url{https://github.com/JeffSackmann/tennis_slam_pointbypoint}.

\section*{Acknowledgments} This work was supported by JST SPRING (Grant Number JPMJSP2108). We would like to thank Jeff Sackmann for making the tennis point-by-point data publicly available through his GitHub repository, which made this research possible. We are grateful to the members of the Sports Biomechanics Lab at the University of Tokyo for their support and valuable discussions. Finally, we thank Tatsuya Kubo for the many stimulating discussions and his friendship, which greatly enriched this work.

\section*{Author Contributions} M.Y. conceived and designed the study, performed all analyses, and wrote the manuscript. 
Y.K. provided feedback on the manuscript and contributed to the interpretation of results. 
D.T. supervised the project and provided conceptual guidance and critical revisions. 
All authors discussed the results and approved the final manuscript.

\section*{Competing Interests} The authors declare no competing interests.

\bibliographystyle{unsrt}  
\bibliography{pnas-references}  

\end{document}